\documentclass[fleqn,twoside]{article}
\usepackage[headings]{espcrc2}

\readRCS
$Id: espcrc2.tex,v 1.2 2004/02/24 11:22:11 spepping Exp $
\usepackage{graphicx}
\usepackage[figuresright]{rotating}

\newcommand{\etal}{\it et al.\rm}
\newcommand{\rt}{\rightarrow}

\newcommand{\AmS}{{\protect\the\textfont2
  A\kern-.1667em\lower.5ex\hbox{M}\kern-.125emS}}

\title{Charmonium Review}

\author{Frederick A. Harris\address[MCSD]{Dept. of Physics and Astronomy, \\        The University of Hawaii, \\ 
        Honolulu, HI 96822, USA}%
        \thanks{Supported by the US Dept. Of Energy under Grant DE-FG03-94ER40833}        
}

\runtitle{2-column format camera-ready paper in \LaTeX}
\runauthor{F. A. Harris}

\begin{document}

\begin{abstract}
During the last few years there has been a renaissance in charm and
charmonium spectroscopy with higher precision measurements at the
$\psi^{'}$ and $\psi(3770)$ coming from BESII and CLEOc and many new
discoveries coming from B-factories.  In this paper, I will review the
status of $\psi(3770)$ and below.

\vspace{1pc}
\end{abstract}

\maketitle

\section{Introduction}
An admonishment says ``May you live in exciting times''.  It turns out
that for charm and charmonium spectroscopy, we are.  Only 10
$C\bar{C}$ resonances were discovered from 1974 to 1977; none were
discovered from 1978 to 2002. However from 2002 to 2005, seven new
$C\bar{C}$ resonances were discovered by Belle, BaBar, CLEOc, CDF, and
D0.  I will review the $\psi(3770)$ and below, where there has been
much progress from BES, CLEOc, B-factories, etc.  Brian Petersen will
cover the new states: X, Y, Z, etc.

\section{Old but new states} 
\subsection{\boldmath $\eta_c^{'}$}
Prior to 2002, there was an unconfirmed candidate for the $\eta_c^{'}$
by the Crystal Ball experiment~\cite{crystal-ball} at a mass of 3594
$\pm$ 5 MeV/c$^2$. In 2002, Belle observed clear peaks in the $X$ mass
distribution in $B\rt KX, X \rt K_SK\pi$ at the $\eta_C$, the
$J/\psi$, and at a mass of 3654 $\pm$ 10 MeV/c$^2$~\cite{etacp}.
CLEO~\cite{cleo_etacp} and BaBar~\cite{babar_etacp} quickly confirmed
the higher mass value in $\gamma \gamma \to K_SK\pi$ with mass
measurements of 3642 MeV/$c^2$ and 3633 MeV/$c^2$, respectively.
Belle also found a peak in $e^+ e^- \rt J/\psi X$ at $M_X = 3630$
MeV/$c^2$~\cite{belle2}

Combining the results, excluding Crystal Ball, yields a mass
of $M_{avg}$ = 3637 $\pm$ 4 MeV$/c^2$, and hyperfine splittings of 
$\Delta M(1S) = M_{J/\psi}-M_{\eta_c} = 117 \pm 1$ MeV/$c^2$ and
$\Delta M(2S) = M_{\psi(2S)}-M_{\eta_c^{'}}= 49 \pm 4$ MeV/$c^2$. 
The higher mass is more consistent
with lattice calculations (LQCD)
and potential models~\cite{consistency}.

\subsection{\boldmath $h_c$}
The $h_c$ or $^1P_1 C\bar{C}$ state has $J^{PC} = 1^{+-}$.  This state
is important to learn more about the hyperfine (spin-spin) interaction
of P wave states. It is expected to have a mass near the
center of gravity of the $^3P_1$ states $m_{h_c} = m_{c. of g.} =
3525.31 \pm 0.07 ~$MeV/$c^2$, to be narrow ($\Gamma
< 1$ MeV/$c^2$), and to decay to $\eta_c \gamma$.

In 1992, E760 using 16 pb$^{-1}$ of $p\bar{p}$ data observed a
structure near 3526 MeV/$c^2$ in $p\bar{p} \rt J/\psi \pi^0$~\cite{E760}.
The successor experiment E835 using 113 pb$^{-1}$
of data was unable to confirm this peak!  However
E835 also searched for $p\bar{p} \rt h_C \rt \gamma \eta_C, \eta_C \rt
\gamma \gamma$ and found a signal at $ M = 3525.8 \pm 0.2 \pm 0.2$
MeV/$c^2$ for $\Gamma = 0.5$ to 1 MeV/$c^2$~\cite{E835}.

CLEOc quickly substantiated this with 
evidence for $h_C$ production from $e^+ e^- \rt \psi(2S) \rt \pi^0
h_C \rt 3 \gamma \eta_C$  at CESR~\cite{cleohc} with a sample  of 
$3 \times 10^6$ $\psi(2S)$ events. Using an inclusive analysis where they
measured the mass recoiling from the $\pi^0$,
they obtained $M(h_C) =  3524.9 \pm 0.7 \pm 0.4$ MeV/$c^2$.
From an exclusive analysis, where they measured
$h_C$ decays to $K_S K\pi$, $K_L K\pi$, $KK\pi\pi$, $\pi\pi\pi\pi$,
$KK\pi^0$, and $\pi\pi\eta$, they
obtained
$M(h_C) = 3523.6 \pm 0.9 \pm 0.5$ MeV/$c^2$.  The consistency between
the two measurements was good giving an overall  
$ M(h_C) = 3524.4 \pm 0.6 \pm 0.4$ MeV/$c^2$, and a product branching fraction
$B(\psi(2S) \rt \pi^0 h_C)B(h_C \rt \gamma \eta_C) =
         (4.0 \pm 0.6 \pm 0.4) \times 10^{-4}$, in agreement with
a PQCD prediction of 
$B = (1.9 - 5.8) \times 10^{-4}$~\cite{kuang}.
The mass splitting
$\Delta M_{hf} =~ <M(^3P_J)> - M(^1P_1)$ 
            $ = 1.0 \pm 0.6 \pm 0.4 $MeV/$c^2$
agrees with expectations ($\approx$0), but the sign and difference are not yet
well enough determined to provide a real test.

Now the
charmonium family below the $\psi(3770)$ is complete, and the mass
values can be used
in potential models to predict masses of higher states.

\section{\boldmath $\psi(2S)$ radiative and hadronic transitions}
\subsection{\boldmath $\psi(2S)$ radiative transitions}
CLEOc with its CsI calorimeter ($\Delta E/E = 5.0$ \% at 100 MeV)
allows a good measurement of the inclusive
$\gamma$ spectrum. Using $3 \times  10^6$ $\psi(2S)$ events, they measured
        $\psi(2S) \rt \gamma \chi_{cJ}$ (E1 transitions) and 
        $\psi(2S) \rt \gamma \eta_c$ (M1 transition)~\cite{CLEOrad}.
Results are shown in Table~\ref{table1}.
Note the big change from PDG04~\cite{PDG04} for $\psi(2S) \rt \gamma
\chi_{c2}$; this will affect $\chi_{c2}$ branching fractions!
The combined transistions, 
$\psi(2S) \rt \gamma \chi_{cJ}, \chi_{cJ} \rt \gamma J/\psi, J/\psi \rt
\mu^+ \mu^-, e^+ e^-$ have also been measured by 
BESII~\cite{BESprod} and CLEOc~\cite{CLEOprod}, and the
product branching fractions are shown in Table~\ref{table2}.
Branching fractions for $\chi_{CJ} \to \gamma J/\psi$ are given in
Table~\ref{table3}.  BESII is calculated using the CLEOc branching
fractions for $\psi(2S) \rt \gamma \chi_{cJ}$ from Table~\ref{table1}
and the BESII results in Table~\ref{table2}.

\begin{table}[htb]
\caption{\label{table1} Radiative decay branching fractions.}

\begin{tabular}{lcc}
\hline
Decay       & PDG04 \% & CLEOc \%~\cite{CLEOrad} \\
\hline
$\psi(2S) \rt \gamma \chi_{c0}$ & $8.6 \pm 0.7 $ & $9.22 \pm 0.47$  \\
$\psi(2S) \rt \gamma \chi_{c1}$ & $8.4 \pm 0.8 $ & $9.07 \pm 0.55$  \\
$\psi(2S) \rt \gamma \chi_{c2}$ & $6.4 \pm 0.6 $ & $9.33 \pm 0.63$  \\
$\psi(2S) \rt \gamma \eta_c$    & $0.28 \pm 0.08 $ & $0.32 \pm 0.07$  \\
\hline
\end{tabular}\\[2pt]
\end{table}

\begin{table}[htb]
\begin{footnotesize}
\caption{Product branching fractions: (BESII~\cite{BESprod}, CLEOc~\cite{CLEOprod})}
\label{table2}
\begin{tabular}{lccc}
\hline
Decay:     & PDG04  & BESII  & CLEOc  \\
$\psi(2S) \rt$    & (\%)   &  (\%)  &  (\%)  \\
\hline
$\gamma \chi_{c0} \rt \gamma \gamma
J/\psi$ & 0.101 $\pm 0.012 $ & -- & $0.14 \pm 0.02$\\
$\gamma \chi_{c1} \rt \gamma \gamma
J/\psi$ & $2.67 \pm 0.15 $ & $2.81 \pm 0.24$ & $3.44 \pm 0.14$ \\
$\gamma \chi_{c2}  \rt \gamma \gamma
J/\psi$ & $1.30 \pm 0.08 $ & $1.62 \pm 0.13$ & $1.85 \pm 0.08$ \\ \hline
\end{tabular}\\[2pt]
\end{footnotesize}
\end{table}

\begin{table}[htb]
\caption{$\chi_CJ \rt \gamma J/\psi$ branching fractions. BESII calculated
  using the BES results of Table~\ref{table2} and the CLEOc results of Table~\ref{table1}.}
\label{table3}
\begin{tabular}{lccc}
\hline
Decay       & PDG04  & BESII  & CLEOc~\cite{CLEOprod}  \\
            &  (\%)  &  (\%)  &  (\%) \\
\hline
$\chi_{c0}$ & $1.18 \pm 0.14 $ & -- & $2.0 \pm 0.3$\\
$\chi_{c1}$ & $31.6\pm 3.3$  & $31.0 \pm 3.2$ & $37.9 \pm 2.2$ \\
$\chi_{c2}$ & $20.2 \pm 1.7 $ & $17.4 \pm 1.8$ & $19.9 \pm 1.3$
 \\ \hline
\end{tabular}\\[2pt]
\end{table}

\subsection{\boldmath $\psi(2S)$ hadronic transitions}
The $\psi(2S) \rt \gamma \chi_{cJ}, \chi_{cJ} \rt \gamma J/\psi,
J/\psi \rt \mu^+ \mu^-, e^+ e^-$ decays can also be used to measure
the processes $\psi(2S) \rt \pi^0 J/\psi$ and $\eta J/\psi$.  These
and $\psi(2S) \rt \pi \pi J/\psi$ results are shown in
Table~\ref{table4}~\cite{BESprod,CLEOprod,BESII,BESIII}.  Note that
isospin is conserved in the CLEOc $\pi^+ \pi^- J/\psi$ to $\pi^0 \pi^0
J/\psi$ ratio.
Using CLEOc + BESII, we determine
$$R = \frac{\Gamma(\psi(2S) \rt \pi^0 J/\psi)}{\Gamma(\psi(2S) \rt \eta
  J/\psi)} = 0.042 \pm 0.004$$
$R$ is much larger than expected using PCAC~\cite{miller} and may
  indicate mixing between $\pi^0$, $\eta$, and $\eta^{'}$~\cite{kroll}.

\begin{table*}[htb]
\caption{Hadronic transitions: $\psi(2S)$ Branching fractions.}
\label{table4}
\begin{tabular}{lccc}
\hline
Decay     & PDG04 & BES  & CLEOc~\cite{CLEOprod}  \\
\hline
$\pi^0 J/\psi$        & $0.10 \pm 0.02$ \% & $0.14 \pm 0.01 \pm
0.01$ \% ~\cite{BESprod} & $0.13 \pm 0.01 \pm 0.01$ \% \\
$\eta J/\psi$         & $3.16 \pm 0.22$ \%& $2.98 \pm 0.09 \pm 0.23$
\% ~\cite{BESprod} & $3.25 \pm 0.06 \pm 0.11$ \%\\
$\pi^+ \pi^- J/\psi$  & $31.7 \pm 1.1 $ \%& $32.3 \pm 1.4$ \%  ~\cite{BESII} &
$33.54 \pm  0.14 \pm 1.10$ \% \\
$\pi^0 \pi^0 J/\psi$ & $18.8 \pm 1.2$ \% & -- & $16.52 \pm 0.14 \pm
0.58$ \% \\
$\frac{\pi^+ \pi^- J/\psi}{\pi^0 \pi^0 J/\psi}$& $1.69 \pm 0.12$ &
$1.75 \pm 0.03 \pm 0.08 $~\cite{BESIII} & $2.03 \pm 0.04$\\
 \\ \hline
\end{tabular}\\[2pt]
\end{table*}

\section{Hadronic decays of charmonium}

\begin{figure}[htbp]
  \centering
\vspace{-10pt}
  \includegraphics[width=0.46\textwidth,height=0.25\textheight]{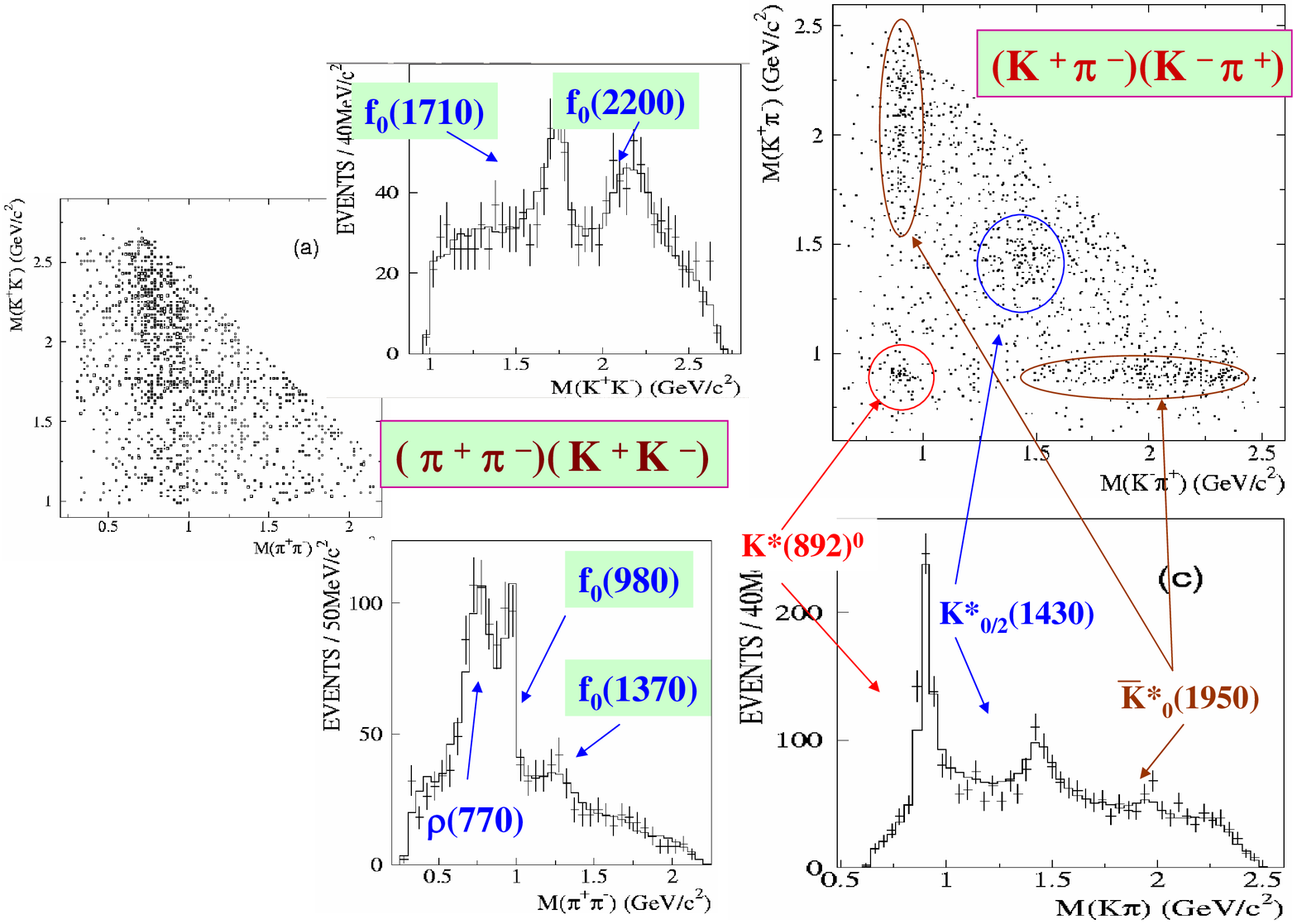}
\vspace{-25pt}
  \caption{Dalitz plots,
projections, and the result of a partial wave analysis fit for
the decay $\chi_{C0} \rt \pi^+ \pi^- K^+ K^-$.}
  \label{chic0pwa}
\end{figure}

Decays of $J/\psi$, $\eta_C$, $\chi_{CJ}$, and $\psi(2S)$ with
definite J and I are ideal to study meson and baryon spectroscopy. In
particular, radiative decays of $J/\psi$ are ideal for glueball searches
~\cite{chan}.  As an example, Fig.~\ref{chic0pwa} shows Dalitz plots,
projections, and the result of a partial wave analysis fit for the
decay $\chi_{C0} \rt \pi^+ \pi^- K^+ K^-$.  The Dalitz plots show rich
structure, and these decays are ideal for studying scalar
states~\cite{chictopipikk}.
 
The pQCD 12\% rule~\cite{politzer,derujula} states that single
$J/\psi$ and $\psi(2S)$ hadronic decays to final state $X$ proceed via
the annihilation of the $C\bar{C}$ pair into three gluons or a virtual
photon, and the decay rate should be determined by the wave function
at the origin squared ($|\psi(0)|^2$), which is measured by the decay
rate into leptons, and therefore
\begin{small}
$$Q_h = \frac{B(\psi(2S)\rt X)}{B(J/\psi \rt X)} =
                \frac{B(\psi(2S)\rt e^+ e^-)}{B(J/\psi \rt e^+e^-)}
                \sim 12 \%$$
\end{small}
MARK-II found that a number of decays obeyed this rule but that it was
                badly violated for $VP$ decays to $\rho \pi$ and $K^*
                K$, the so called $\rho \pi$ puzzle~\cite{franklin}. 
The suppression was confirmed by BESI with higher sensitivity, and BESI
also found the VT mode to be suppressed~\cite{VT}.

\begin{figure}[htbp]
  \centering
\vspace{-10pt}
  \includegraphics[width=0.51\textwidth,height=0.25\textheight]{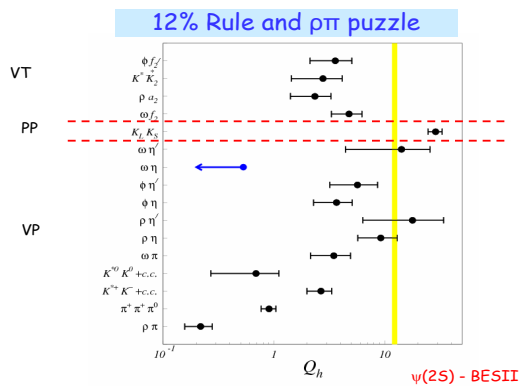}
\vspace{-25pt}
  \caption{Measurements of $Q_h$ for some two body decays by BESII.}
  \label{fig2}
\end{figure}

There have many attempts at theoretical explanations~\cite{theory}.
Together BESII, CLEOc, and BaBar have all made many new $J/\psi$ and
$\psi(2S)$ branching
fraction measurements~\cite{PDG06}.
A summary of a few new $Q_h$ values from BESII is shown in Fig.~\ref{fig2}.
There is no obvious rule to categorize the suppressed, the enhanced,
and the normal decay modes of the $J/\psi$ and $\psi(2 S)$.
Hopefully the many new measurements will help in understanding this problem.

\section{\boldmath $\psi(3770)$}
The $\psi(3770)$ is just above $D \bar{D}$ threshold so it decays
mostly to correlated $D \bar{D}$ pairs. Its importance for charm
physics has been stressed by many speakers.  BESII has 34 pb$^{-1}$ at
and around the $\psi(3770)$, and CLEOc has 281 pb$^{-1}$ at the
$\psi(3770)$.  These samples not only allow precision charm decay
measurements, they also better our understanding of the $\psi(3770)$.

\begin{figure}[htbp]
  \centering\vspace{-20pt}
  \includegraphics[width=0.50\textwidth,height=0.25\textheight]{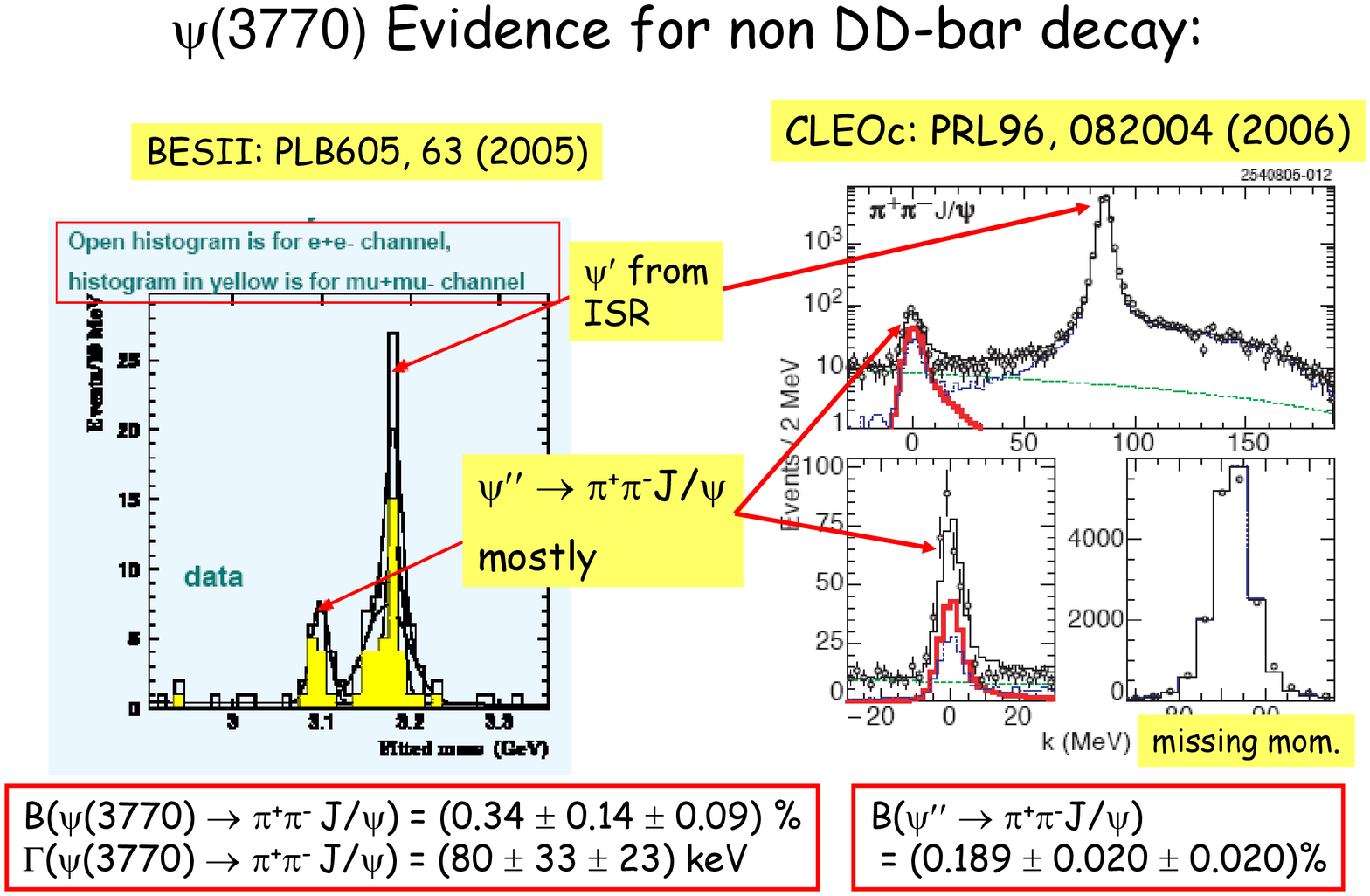}
\vspace{-25pt}
  \caption{Evidence for non $D \bar{D}$ decay through  $\psi(3770) \rt \pi^+
\pi^- J/\psi$ by BESII and CLEOc.}
  \label{fignondd}
\end{figure}

The $\psi(3770)$ is thought to be a mixture of $S$ and $D$ wave (mostly
$D$), but
since the $\psi(3770)$ is above DD-bar threshold it is expected to
decay mostly to $D \bar{D}$.
BESII found evidence (see Fig.~\ref{fignondd}) for non  $D \bar{D}$ decay in $\psi(3770) \rt \pi^+
\pi^- J/\psi$~\cite{nondd} with a branching fraction $B(\psi(3770) \rt \pi^+
\pi^- J/\psi) = (0.34 \pm 0.14 \pm 0.09) \%$ and a width of $\Gamma(\psi(3770) \rt \pi^+
\pi^- J/\psi) = (80 \pm 33 \pm 23)$ keV, to be compared to a prediction
of 26 to 147 keV~\cite{kuang}.  CLEOc with a larger data
sample confirmed this with $B(\psi(3770) \rt \pi^+
\pi^- J/\psi) = (0.189 \pm 0.020 \pm 0.020) \%$~\cite{cleonondd}.

CLEOc has also found evidence for non  $D \bar{D}$ decays in the hadronic
transitions, $\psi(3770) \rt \pi^0 \pi^0 J/\psi$ and  $\psi(3770) \rt
\eta J/\psi$~\cite{cleonondd} and in radiative decays $\psi(3770) \rt
\gamma \chi_{CJ}$~\cite{cleochicj1,cleochicj2}.  However, they have
found that hadronic decays at the $\psi(3770)$ are mostly consistent with
continuum production~\cite{CLEOhadronic}. 

\section{Future}
We can expect further progress from B factories, BESII, and CLEOc.
However, BaBar
and CLEOc will stop running in 2008.  BEPCII with a design luminosity
of $1 \times 10^{33}$ cm$^{-2}$ s$^{-1}$ and a brand new BESIII
detector will start commissioning in summer of 2007~\cite{fahbes3}.
With B factories, CLEOc, and BESIII, the future of charm and
charmonium physics is very bright.

\begin{footnotesize}

\end{footnotesize}

\end{document}